\documentclass[conference]{IEEEtran}
\usepackage{blindtext, graphicx}%, subfigure}
\usepackage{amsmath, amsthm, amssymb, algorithmic, array, mathtools}
\usepackage{breqn}
\usepackage{empheq}
\usepackage{verbatim}
\usepackage[noadjust]{cite}
\usepackage{relsize}
\usepackage{epstopdf}

\usepackage{booktabs}% http://ctan.org/pkg/booktabs

\IEEEoverridecommandlockouts

\begin{document}
\bstctlcite{IEEEexample:BSTcontrol}
\title{\huge Learning How to Communicate in the Internet of Things: Finite Resources and Heterogeneity \vspace{-.2cm}}

\author{\IEEEauthorblockN{Taehyeun Park$^1$, Nof Abuzainab$^1$, and Walid Saad$^1$ \thanks{This research was supported by the U.S. Office of Naval Research (ONR) under Grant N00014-15-1-2709.}}

\IEEEauthorblockA{\small $^1$Wireless@VT, Bradley Department of Electrical and Computer Engineering, Virginia Tech, Blacksburg, VA, USA,\\ Emails:\{taehyeun, nof, walids\}@vt.edu \vspace{-.5cm}}}

\maketitle

\begin{abstract}
For a seamless deployment of the Internet of Things (IoT), there is a need for self-organizing solutions to overcome key IoT challenges that include data processing, resource management, coexistence with existing wireless networks, and improved IoT-wide event detection. One of the most promising solutions to address these challenges is via the use of innovative learning frameworks that will enable the IoT devices to operate autonomously in a dynamic environment. However, developing learning mechanisms for the IoT requires coping with unique IoT properties in terms of resource constraints, heterogeneity, and strict quality-of-service requirements. In this paper, a number of emerging learning frameworks suitable for IoT applications are presented. In particular, the advantages, limitations, IoT applications, and key results pertaining to machine learning, sequential learning, and reinforcement learning are studied. For each type of learning, the computational complexity, required information, and learning performance are discussed. Then, to handle the heterogeneity of the IoT, a new framework based on the powerful tools of cognitive hierarchy theory is introduced. This framework is shown to efficiently capture the different IoT device types and varying levels of available resources among the IoT devices. In particular, the different resource capabilities of IoT devices are mapped to different levels of rationality in cognitive hierarchy theory, thus enabling the IoT devices to use different learning frameworks depending on their available resources. Finally, key results on the use of cognitive hierarchy theory in the IoT are presented.
\end{abstract}

\section{Introduction} %% add citations
\indent The Internet of Things (IoT) is a complex ecosystem that will interconnect smartphones, tablets, machine type devices (MTDs), people, and mundane objects into a large-scale interconnected network \cite{saad, ml1}. Such a pervasive ecosystem will deliver innovative applications and services including drone-based services \cite{moha1, moha2, drone1}, smart grid features \cite{saad2, smartgrid, smartgrid2}, and new healthcare applications \cite{ml1}. An effective delivery of such IoT services requires a reliable wireless infrastructure that can enable communications within the heterogeneous IoT environment. Such wireless systems can range from the popular wireless cellular network to the emerging long-range wide-area network (LoRaWAN) \cite{lora}. However, there are many challenges in incorporating the IoT devices into wireless systems that include processing massive volumes of data, coexistence with existing systems, stringent resource constraints, new quality-of-service (QoS) requirements, and heterogeneity in traffic patterns and device types \cite{saad}.\\
\indent These challenges motivate the deployment of novel resource management mechanisms using which the scarce wireless resources, such as power and frequency, can be properly managed while being cognizant of the unique nature of communications in the IoT \cite{saad}. IoT devices, such as MTDs, will inherently exhibit stringent resource constraints in terms of memory, energy, and computation. The coexistence of human-to-human (H2H) communications with machine-to-machine (M2M) communications must also be properly managed so as not to jeopardize the QoS of existing H2H services while also meeting the QoS of new IoT services. Moreover, while conventional H2H services typically seek to maximize data rates, IoT services may be more concerned with reliability or latency \cite{lat3, lat4}, thus bringing forward new QoS challenges for resource management. Furthermore, unlike in conventional systems, the IoT will have a communication bottleneck in the uplink, because typical IoT applications, such as smart sensing or wearable communications, will mostly be uploading collected data to a base station (BS) rather than downloading information from the BS \cite{saad}.\\
\indent To develop resource management mechanisms tailored to the IoT, different approaches, such as optimization theory \cite{manage1, moha3, opt1} or game theory \cite{manage3}, have been applied. In this regard, optimization techniques are often centralized, which can incur unnecessary overhead and will thus require constant, energy-consuming communication between the energy limited IoT devices and BS. Moreover, most optimization frameworks cannot naturally handle the heterogeneity of IoT devices without significant additional overhead. One recently proposed solution was via the use of multi-objective optimization \cite{moo}, however, such an approach is often suboptimal and may not scale well as the network size increases. One other promising resource management approach for the IoT is via the use of game-theoretic solutions \cite{manage3}. However, to achieve such game-theoretic solutions, there is a need for distributed algorithms that can converge to an equilibrium in a timely manner. Indeed, the design of such algorithms is essential to practically deploy any game-theoretic construct within a real-world wireless system, such as the IoT.\\
\indent One effective approach to overcome the aforementioned challenges and enable self-organizing operation of the IoT is via the use of \emph{learning frameworks}. Learning will allow the IoT devices to adapt to their dynamic environment, adjust their performance parameters, learn and process statistical information gathered from the users and other IoT devices, and optimize the overall system performance \cite{mlbook, learning1, cover, rlbook}. For the IoT, learning frameworks are particularly useful for managing limited resources and handling the aforementioned IoT properties \cite{learning2, tae, fmrl}. Indeed, learning techniques are inherently distributed, thus enabling the IoT devices to operate without draining their limited energy by communicating constantly with a centralized controller. Furthermore, if properly designed, learning can be computationally simple, and the IoT devices can learn their resource management parameters in a timely manner to meet QoS requirements. Moreover, learning enables the IoT devices to adapt to the dynamic environment autonomously with minimal human intervention. Despite the promising outlook of using learning techniques in the IoT, it is necessary to carefully design such techniques to account for various unique IoT properties, such as the resource limitations of the devices and the heterogeneity of the system.\\
\indent To allow IoT devices with heterogeneous resource capabilities to use different suitable learning frameworks, \emph{cognitive hierarchy theory} \cite{Camerer_acognitive, CHIoT} can be used to capture the heterogeneity in available resources, such as computation, memory, and energy, in a hierarchy of devices ranging from sensors to smartphones. Cognitive hierarchy theory is a mathematical framework that captures different levels of agent rationality, which, in the IoT, correspond to different levels of resources available to the IoT devices. Since IoT devices with more resources available will be able to perform more sophisticated optimization, the use of cognitive hierarchy models will allow these devices to use different learning techniques depending on their resources and thus provide more realistic model of IoT.\\
\indent The main contribution of this article is to provide a comprehensive overview on the use of learning techniques within the scope of the IoT. Despite some recent surveys on the challenges of wireless communications in the IoT, such as in \cite{survey1} and \cite{survey2}, none of these works have investigated the potential of developing learning techniques for the IoT. As such, to our best knowledge, this will be the first comprehensive tutorial on this topic. First, we discuss the various types of learning frameworks that are suitable to address a number of key challenges of the IoT. Then, for each learning approach, we outline the basic concepts, main challenges, potential applications, and key results. Subsequently, to handle system heterogeneity, we introduce the basics of cognitive hierarchy theory and its application to the IoT. Then, we draw the key connections between cognitive hierarchy theory and the different classes of learning algorithms overviewed. We conclude by summarizing the potential of learning in the IoT.\\
\indent The rest of this paper is organized as follows. In Section II, we present the motivations and challenges of applying learning frameworks in the IoT. Section III discusses the advantages, limitations, and applications of the learning frameworks within the scope of the IoT. In Section IV, we introduce cognitive hierarchy theory and analyze its implementation with learning frameworks in the IoT. Finally, we summarize our key findings in Section V.
%%%%%%%%%%%%%%%%%%%%%%%%%%%%%%%%%%%%%%%%%%%%%%%%%%%%%%%%%%%%%%%%%%%%%%%%%%%%%%
\section{Learning in the IoT: Motivation and Challenges}
\subsection{Motivation}%%%%%%%%%%%%%%%%%%%%%%%%%%%%%%%%%%%%%%%%%%%%%%%%%%%%%%%%%%%%%%%%%%%%%%%%%%%%%%
\indent The sheer scale of the IoT makes it impractical to manage the devices manually, and, hence, it is essential to operate the devices in a self-organizing manner, with minimal human intervention. Moreover, the devices will be deployed in various environments depending on their applications. Examples include deployments in an urban area with many diverse devices, a distant forest with few sensors for environmental monitoring, an indoor environment for home automation, or a rural area with devices spread across a large field for agricultural applications. Furthermore, the deployment environment for the IoT will be highly dynamic due to unpredictable events, such as weather conditions, power outages, or medical emergencies. Such unpredictable events will trigger uplink transmission links that must be serviced with low latency and high reliability due to their urgent nature \cite{tae}. However, satisfying the QoS requirements of such uplink transmissions is challenging, because existing wireless networks have been designed primarily for H2H communications and the sheer scale of the IoT will strain their resources, thus requiring new approaches for enhancing network operation.\\
\indent The adaption of learning techniques will provide an effective solution for enabling the IoT devices to adapt to dynamic environments, manage their limited resources, and satisfy the stringent QoS requirements. To manage its resource consumption, an IoT device can learn about its environment, devices, users, and their usage patterns. For instance, MTDs can learn about their usage patterns in terms of peak workload intervals to adjust their status, such as active or sleep, to minimize the energy consumption \cite{learning2}. Moreover, learning can be used to coordinate the usage of limited radio resources among the IoT devices and human type devices (HTDs). For example, unpredictable events may trigger uplink communication links with stringent QoS requirements. Here, by using a proper learning framework, the uplink communication resources can be allocated for the IoT devices that are sending the data pertaining to the unpredictable events in a way to minimize the effects, such as throughput reduction or increased latency, on the normal uplink communications of IoT. Furthermore, learning can also enable the IoT to process and learn from the massive volume of data collected from the IoT devices. Indeed, learning is a key enabler for big data analytics that will be used to process the data related to the IoT, its human users, and its environment. Therefore, learning techniques in the IoT can be used to extend the lifetime of IoT devices, to mitigate the uplink communication bottleneck, to promote a coexistence of IoT and pre-existing systems, and to process the massive data from IoT.
\subsection{Challenges}%%%%%%%%%%%%%%%%%%%%%%%%%%%%%%%%%%%%%%%%%%%%%%%%%%%%%%%%%%%%%%%
\indent To effectively use learning for solving IoT-centric problems, several challenges that stem from the unique characteristics and requirements related to the IoT and its devices must be met, as described next.
\begin{itemize}
\item A key property of low-cost, low-capability MTDs is \emph{low computational capability} as discussed in \cite{saad} and  \cite{learning1}. However, existing learning frameworks, such as decision trees \cite{learning1} or reinforcement learning \cite{moha3} and \cite{luca}, can be computationally complex to be adapted by the resource-constrained MTDs.
\item \emph{Energy limited} MTDs are expected to have \emph{extended lifetime} with \emph{minimal human intervention} as pointed out in \cite{ml1, manage1}, and \cite{ratasuk2}. Since a centralized learning framework will require a constant communication with the BS and potentially cause an excessive energy consumption, learning for the IoT devices must be distributed. Moreover, the MTDs may be deployed in adverse locations in which the BS is not always accessible to deploy a centralized framework. Therefore, a distributed learning technique is needed to lower energy consumption and enable the IoT devices to be self-organizing.
\item To implement distributed learning methods in the IoT, the necessary information, such as the actions of other IoT devices, may need to be collected via resource consuming M2M communication links. However, with a lack of radio resources in the IoT and energy constraints of the IoT devices as pointed out in \cite{ml1, saad}, and \cite{learning1}, a frequent M2M communication to collect the necessary information for learning is not viable for the IoT devices, and thus they may have \emph{limited information available}. Therefore, any deployed learning framework must be able to accurately adjust the performance parameters with limited information. 
\item Certain IoT applications, such as industrial control or health monitoring, are of critical importance, and thus such applications require \emph{ultra-reliable, low-latency communication} as discussed in \cite{lat3, lat4}, and \cite{lat2}. To satisfy such QoS requirements, the IoT devices must quickly learn to adjust their performance parameters. Therefore, the time necessary to converge to a steady state must be short.
\item The IoT may be deployed using an existing wireless network, and thus \emph{M2M communication will coexist with pre-existing communication links} as pointed out in \cite{saad} and \cite{survey2}. In particular, the IoT is likely to be deployed over a wireless cellular network in which M2M communications will have to coexist with H2H communications. For harmonious coexistence, the learning techniques for the IoT must consider both existing traffic and the new traffic potentially stemming from the IoT. This is due to the fact that coexisting wireless networks will inevitably impact each other due to factors, such as interference, and it is necessary to not disrupt the existing network and its services.
\end{itemize}

\indent To address the aforementioned challenges, next, we introduce different classes of learning techniques, while providing a key overview on their applicability to the IoT. 
\section{Classification of Learning Frameworks for the IoT}%%%%%%%%%%%%%%%%%%%%%%%%%%%%%%%%%%%%%%%%%%%%%%%%%%%%%%%%%%%%%%%%%%%%%%%%%%%%%%
\indent Learning has been developed for many disciplines ranging from socioeconomic to mathematical modeling which has led to a variety of learning frameworks having different properties, objectives, and capabilities, such as in \cite{mlbook, learning1, tae, fmrl, rlbook}, and \cite{fm1}. For the IoT, there is no single learning framework that can be used to overcome all of its aforementioned challenges, and, thus, it is necessary to characterize the key learning frameworks that can be useful for various IoT applications. For each such framework, one needs to identify the main advantages and limitations. For example, a learning framework that is computationally simple but requires a long time to converge is suitable for the delay-tolerant IoT applications involving simple sensors, however, this will not be viable for critical applications requiring low latency.\\
\indent To this end, we propose to classify the rich literature on learning into three classes of learning frameworks that include: a) machine learning, b) sequential learning, and c) reinforcement learning. This classification will allow us to discuss the main benefits of implementing each class in the IoT and to determine the learning frameworks that are suitable for a given aspect of the IoT. Indeed, machine learning, sequential learning, and reinforcement learning are proven to be particularly useful to address problems pertaining to big data analytics, enhanced event detection, and resource management, respectively, as in \cite{learning1, tae}, and \cite{fmrl}. For each class of learning, we will identify the main advantages, challenges, IoT applications, and key results. In addition, we will identify their inherent properties, in terms of computations, information requirement, and capabilities. Here, we note that, although learning has been very popular for wireless networks \cite{luca}, there has not been yet any comprehensive overview on applicability and limitations of learning in the IoT.
\subsection{Machine Learning}%%%%%%%%%%%%%%%%%%%%%%%%%%%%%%%%%%%%%%%%%%%%%%%%%%%%%%%%%%%%%%%%%%%%%%%%%%
\subsubsection{Key Concepts of Machine Learning}%%%%%%%%%%%%%%%%%%%%%%%%%%%%%%%%%%%%%%%%%%%%%%%%%%%%%%%%%%%%%%%%%%%%%%%%%%%%%%
\indent Machine learning (ML) techniques were originally developed for allowing computers to autonomously learn information from existing data sets and, subsequently, build suitable models to make a decision on future actions and behaviors as discussed in \cite{mlbook} and \cite{learning1}. ML techniques are typically categorized into supervised and unsupervised learning \cite{mlbook}. Supervised learning requires a labeled training data, while unsupervised learning, such as clustering, does not require labels. However, unsupervised learning is more computationally complex than supervised learning. Furthermore, for certain ML techniques, there may be specific requirements for the training data set. For instance, decision trees require the data set to be linearly separable \cite{mlbook}.\\
\indent Using the training data set, ML techniques can build regression models to determine the relationship between the variables, divide the feature space to classify the unlabeled data points, cluster the data points to divide into different groups, or lower the dimensionality of feature space by removing the correlated features as discussed in \cite{mlbook, ml1}, and \cite{ml2}. However, ML techniques may be computationally complex and implicitly require a central entity with significantly high computational capabilities. Therefore, ML techniques are often centralized.
\subsubsection{Limitations and Opportunities of ML for the IoT}%%%%%%%%%%%%%%%%%%%%%%%%%%%%%%%%%%%%%%%%%%%%%%%%%%%%%%%%%%%%%%%%%%%%%%%%%%%%%%
\indent The biggest limitation of applying ML in IoT scenarios is that ML requires an extensive data set, such as the information of sensor locations and corresponding sensor measurements \cite{learning1}, for good performance. Such a data set needs to be quickly processed for the IoT devices to learn the environment, but the resource-constrained IoT devices may not be able to store and process the data set given that they have limited resources in terms of computation and memory \cite{ml2}. Furthermore, supervised learning requires a labeled training data set, which can require human intervention to provide the correct labels. This is due to the fact that assuming that the IoT devices are able to label the data points correctly and autonomously would imply that their learning is already done a priori. Therefore, the use of ML must be properly tailored to such IoT features and device limitations.\\
\indent Nonetheless, ML has been recently used in the IoT as a centralized framework as in \cite{learning1, bigdata, bigdata2, pca1, pca1e}. This is because a cloud-based centralized processing unit can be used to implement effective ML schemes. Since ML has a great potential in processing and analyzing the massive data collected in an IoT environment as discussed in \cite{ml1} and \cite{bigdata}, such a cloud-based processing unit will allow the IoT to run the ML techniques for big data analytics purposes \cite{ml1}. The challenges of implementing a cloud-based processing unit for the IoT are designing a scalable wireless architecture with many cooperating wireless access points, developing advanced antenna technologies to boost throughput and reliability, and ensuring the confidentiality and security of data \cite{bigdata}. However, with a cloud-based processing unit, it is possible to determine spatial, temporal, and social correlations of the data traffics to reduce communication overheads, increase energy efficiency, and provide user-oriented services as discussed in \cite{bigdata} and \cite{bigdata2}.
\subsubsection{Applications of ML in the IoT}%%%%%%%%%%%%%%%%%%%%%%%%%%%%%%%%%%%%%%%%%%%%%%%%%%%%%%%%%%
\indent One of the key applications of ML in the IoT is big data analytics as massive data will be accumulated at either BS or local data aggregators \cite{ml1}. The key function of ML in the IoT is to eliminate the correlated information and to reduce the dimensionality of big data in the IoT \cite{learning1}. This will make the data transmission from the data aggregators to BS less costly and the data processing at BS more efficient \cite{learning1}. However, it is necessary to maintain the key information in the data.\\
\indent Principle component analysis (PCA) has been used to intelligently compress the collected data by reducing the dimensionality, while keeping the key features of the data as discussed in \cite{learning1, bigdata, pca1}, and \cite{pca1e}. PCA uses an orthogonal transformation to output a set of linearly uncorrelated data from possibly a correlated input data. PCA has been shown to be effective in reducing the resource consumption by compressing the data. For instance, PCA based query reduces the energy consumption by at least 25\% compared to the normal query \cite{bigdata2}. Therefore, PCA is particularly useful for the data aggregators and the BS for data processing. However, PCA is computationally complex and can require a significant time to process \cite{learning1}. Although the high computational complexity may be feasible for a BS with sufficient resources, it will be challenging to deploy at the level of a data aggregator, which is often a typical IoT device with limited resources.\\
\indent Since data aggregation is one of the popular solutions to mitigate the lack of radio resources in the IoT as discussed in \cite{learning1, pca1} and \cite{pca1e}, it would be worthwhile to investigate using different PCA-based approaches to analyze the tradeoff between computational complexity, data compression, and resource consumption. Moreover, analyzing different PCA-based approaches will help choosing an optimal data aggregation method for a given IoT application. For instance, PCA-based data aggregation with small delay can be used for the low latency IoT applications. Additionally, emerging ML techniques, such as echo state networks, can be useful to develop predictive approaches for resource allocation in the IoT as discussed in \cite{mm}. Finally, ML has been recently shown to be effective in deploying new security techniques for the IoT, such as IoT device authentication \cite{security1} or static malware analysis \cite{security2}.
\subsection{Sequential Learning}%%%%%%%%%%%%%%%%%%%%%%%%%%%%%%%%%%%%%%%%%%%%%%%%%%%%%%%%%%
\subsubsection{Key Concepts of Sequential Learning}%%%%%%%%%%%%%%%%%%%%%%%%%%%%%%%%%%%%%%%%%%%%%%%%%%%%%%%%%%
\indent In an environment with an underlying binary state, a number of autonomous agents can learn what the true underlying state of the environment is by using sequential learning (SL) as discussed in \cite{cover} and \cite{infocas}. Here, such agents are intelligent entities that are capable of learning from a given information, and they would potentially correspond to the IoT devices if SL is implemented in the IoT. Here, the underlying state describes the current state of the IoT environment in which the agents are located, and it corresponds to events of interest, such as a medical status, fire alarm, or other environment-triggered events, within the IoT \cite{tae}. Using SL, the agents learn the state of the system by following a given order while observing the environment and the actions or observations of previous agents in the sequence. Moreover, the agents will eventually converge to a consensus on the true underlying state with repeated hypothesis testing as explained in \cite{cover} and \cite{infocas}. For different agents, the observations of the environment, which are also known as the \emph{private signals}, are independent and equally informative about the underlying state as pointed out in \cite{cover} and \cite{fm1}. Such private signals are typically modeled as independent binary signals whose distribution depends on the underlying state such as in \cite{tae} and \cite{fm1}. Furthermore, a private signal cannot fully reveal the underlying state of the system, which means that the likelihood ratio cannot be zero or infinity \cite{fm1}.

\begin{figure}[t]
	\centering
	\includegraphics[scale = 0.8]{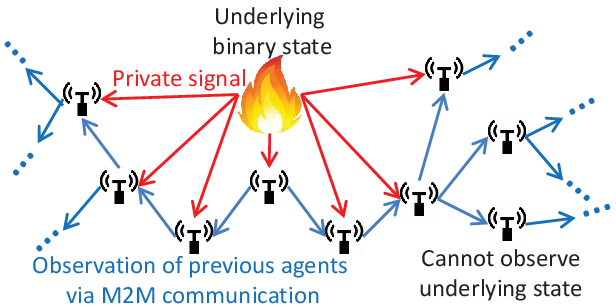}
	\caption{Sequential Learning in the IoT.}
	\vspace{-0.4 cm}\label{fig:slpic}
\end{figure}

\indent Another information necessary for SL is the observation of previous agents and specifically their estimates of the underlying state. Depending on the number of previous estimates required, SL can be classified into \emph{infinite} and \emph{finite memory} as discussed in \cite{cover}, \cite{fm1}, and \cite{infocas}. In infinite memory SL, agents must observe all previous agents in the sequence, and thus the memory of previous estimates grows infinitely. In finite memory SL, agents only observe a fixed number of previous agents, and thus the memory of previous estimates is fixed \cite{tae}. SL with finite memory requires much less information than infinite memory SL at the expense of higher probability of error in estimating the underlying state \cite{tae}.\\
\indent One of the favorable properties for implementing SL approaches in the IoT is that SL can converge to the correct underlying state by only observing two previous agents as discussed in \cite{cover} and \cite{fm1}. Unlike ML, SL does not require an extensive data set, but it relies on the sequential propagation of information among the agents. Moreover, SL can be implemented using distributed mechanisms, while the majority of ML techniques are centralized. Furthermore, SL is more suitable for enhancing event detection, while ML techniques are more suitable for data analytics.
\subsubsection{Limitations and Opportunities of SL for the IoT}%%%%%%%%%%%%%%%%%%%%%%%%%%%%%%%%%%%%%%%%%%%%%%%%%%%%%%%%%%
\indent The majority of information necessary for SL stems from other agents, and, thus, SL must rely on the use of direct M2M communication links as illustrated in Fig. \ref{fig:slpic}, which can consume additional network resources. Moreover, in SL, the IoT devices that are not able to communicate with any other device will not be able to learn. For infinite memory SL, the amount of necessary information will increase indefinitely as learning progresses. This implies that the size of the M2M communication packets may grow infinitely, which can become prohibitive in an IoT environment. However, finite memory SL will only require the packets of fixed size to be transmitted using M2M links, thus making it much more viable for IoT.\\
\indent Another limitation of SL is that it is typically limited to learning a binary state. Although there are practical IoT scenarios that only have two states, such as whether to go to sleep mode to conserve energy or not in minimizing power consumption, many IoT scenarios will involve more than two states, such as multiple levels of transmit power. Furthermore, the private signals in the IoT may not be equally informative as the IoT devices can have different observations of the environment as shown in Fig. \ref{fig:slpic}. However, one can overcome this private signal assumption by introducing additional information about previous IoT devices \cite{tae}.\\
\indent The biggest advantage of applying SL in the IoT is its flexibility with memory requirements. For finite memory SL, the IoT devices can observe different number of previous agents, and the SL will still converge \cite{tae}. Therefore, the IoT devices can optimize the amount of information to use for SL depending on the available resources. However, there is a lower bound on the number of observations of previous agents necessary for the convergence \cite{fm1}. 
\subsubsection{Applications of SL in the IoT}%%%%%%%%%%%%%%%%%%%%%%%%%%%%%%%%%%%%%%%%%%%%%%%%%%%%
\begin{figure}[t]
	\centering
	\includegraphics[scale = 0.6]{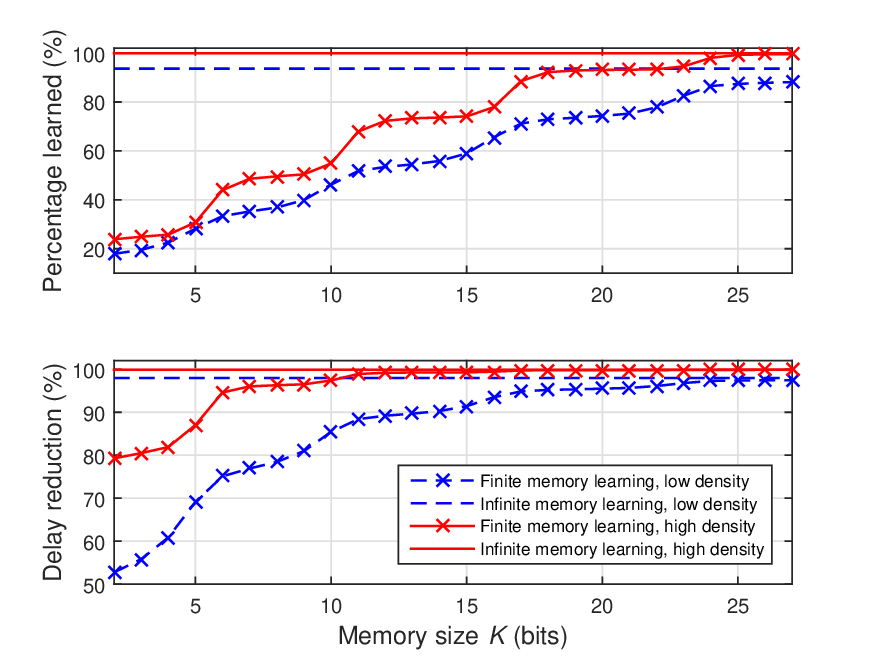}
	\caption{Effectiveness of SL for Different Memory Sizes and Device Densities.}
	\vspace{-0.4 cm}\label{fig:seqlearn}
\end{figure}

\indent SL is particularly useful in enhancing the IoT event detection, and it can be used for distributed resource allocation in the IoT. For instance, there may be urgent events, such as a medical emergency, that trigger uplink communications requiring low latency. However, satisfying the QoS requirements of such communications is not trivial due to the uplink communication bottleneck caused by the IoT devices regularly reporting their data. In such a case, the IoT devices using SL can collectively and autonomously detect the urgent events and then allocate necessary radio resources for the event-triggered communications. Since periodic data reporting to the BS is also important, resource allocation for the event-triggered communications must be done in a way to minimize the amount of resources that need to be re-allocated from the devices that have periodic data to send.\\
\indent Finite memory SL has been shown to be effective in autonomously allocating the radio resources to reduce the delay for the urgent, event-triggered communications in the IoT with limited uplink communication resources in \cite{tae}. Using our proposed approach based on finite memory SL in \cite{tae}, the IoT devices transmitting delay-tolerant messages can collectively reallocate uplink communication resources for the IoT device transmitting urgent messages, while minimally affecting the system throughput and significantly reducing the delay of the urgent message. In \cite{tae}, it is shown that the delay of urgent communication can be further reduced at the cost of bigger memory size for SL and greater throughput reduction of the periodic communications. A key feature of finite memory SL is that the memory size can be optimized to satisfy a given QoS requirements. Fig. \ref{fig:seqlearn}, based on our SL model in \cite{tae}, shows that more devices will learn the correct state of the network and the delay will be reduced more with bigger memory sizes. However, bigger memory sizes will require more radio and energy resources to be consumed to transmit and to process the information. Therefore, the memory size must be chosen appropriately considering the performance improvements and available resources. Furthermore, Fig. \ref{fig:seqlearn} shows that the finite memory SL is more effective with higher device deployment density. This is due to the fact that, in a dense IoT environment, there will be a few devices that cannot communicate with any other device, which prevents such devices from learning. This also shows that SL-based techniques will generally be M2M communication reliant. 

\subsection{Reinforcement Learning}%%%%%%%%%%%%%%%%%%%%%%%%%%%%%%%%%%%%%%%%%%%%%%%%%%%%%%%%%%
\subsubsection{Key Concepts of Reinforcement Learning}%%%%%%%%%%%%%%%%%%%%%%%%%%%%%%%%%%%%%%%%%%
\indent Reinforcement learning (RL) is a learning technique in which a number of agents learn how to act by interacting with the environment as discussed in \cite{learning1} and \cite{rlbook}. An RL algorithm is typically composed of a number of agents with their corresponding sets of actions, an environment with a set of states, a state transition function, an immediate reward function, and an initial observation function \cite{rlbook}. At the beginning of each time period, each agent observes the environment and takes an appropriate action to maximize the immediate or future reward. At the end of each time period, each agent will receive an immediate reward, and the state of the environment will change according to the state transition function. The agents iterate according to this process to learn and to converge to a steady state as explained in \cite{learning1} and \cite{rlbook}.\\
\indent One of the unique features of RL is the use of action-reward combination as a feedback to the agents. For instance, using the future rewards as a feedback, RL can be used to maximize their long-term rewards as pointed out in \cite{learning1, rlbook}, and \cite{fmrl}. Furthermore, the function used for choosing the current action can be computationally simple. For instance, the well-known $Q$-function, which is purely algebraic, can be used \cite{learning1}.\\
\indent Although RL can be computationally simple, it can take significant time to converge to a steady state as discussed in \cite{fmrl} and \cite{rlsurvey}. This is because RL learns by exploring different states. Although RL does not require an extensive training data, it requires its agents to know the state transition function as pointed out in \cite{learning1} and \cite{rlbook}. The slow convergence and the requirement to know the state transition function are key challenges facing the use of RL in an IoT environment, however, the unique design of having actions and rewards between the agents and the environment makes RL very useful to treat a number of IoT problems as mentioned in \cite{rein1, rein2, rein3}.
\subsubsection{Limitations and Opportunities of RL for the IoT}%%%%%%%%%%%%%%%%%%%%%%%%%%%%%%%%%%%%%%%%%%%%%%%%%%%%%%%%%%%
\indent The assumption that the state transition function is known is generally needed for many RL algorithms, however, such an assumption may not hold true in the IoT. Indeed, the environment in which the IoT devices are deployed is subject to unpredictable events and uncertainties, such as abnormal sensor readings due to sudden increase in noise, device failures, and unexpected obstacles preventing normal device operation as discussed in \cite{fmrl} and \cite{rlsurvey}. As such, it is possible for the IoT devices to experience an incomplete state transition function. Moreover, it is impractical to account for all possible scenarios that the IoT devices may face in designing the state transition function. Therefore, there may be hidden or unknown states, which the IoT devices can reach at some point. These hidden states are critical to RL as they hinder with determining the reward maximizing action as pointed out in \cite{fmrl} and \cite{rlsurvey}.\\
\indent The hidden or unknown states in RL can be mitigated by introducing a memory of previous actions \cite{fmrl}. In this case, the IoT devices can combat uncertainties in choosing an action at the cost of higher computational complexity and slower convergence \cite{fmrl}. Therefore, RL may not be directly applicable to low latency IoT applications due to its potentially slow convergence.\\
\indent Although RL may have a slower convergence than SL, it does not require significant interactions between the devices over M2M links, which will save significant amount of energy, and it can be used to find the equilibrium solutions for game-theoretic models \cite{luca}. Moreover, RL does not require the devices to be globally synchronized and to engage in the learning process at the same rate, which is advantageous in a distributed system such as the IoT \cite{learning1}. Furthermore, RL can be computationally simple for low-capability devices in IoT \cite{learning1}. Due to its aforementioned, desirable properties, RL has been one of the most popular learning frameworks in the IoT \cite{rein1, rein2, rein3}.
\subsubsection{Applications of RL in the IoT}%%%%%%%%%%%%%%%%%%%%%%%%%%%%%%%%%%%%%%%%%%%%%%%%%%%%%%%%%%%%%%%%%%%%%%%%%%%%%%%%%%%%
\indent RL can be used for IoT resource management, because the components of RL, such as actions and states, can be easily mapped to the corresponding components of resource management in the IoT \cite{rein1,rein2,rein3}. For example, for power management application, the states can be mapped as energy policies, the actions can be mapped as choosing the next best energy policy, and the rewards can be mapped as amount of energy saved \cite{learning2}. More recently, the use of RL has been investigated in \cite{rein1} as a mean to provide optimized resource management for an IoT system that relies on drones to deliver communications to disaster affected areas using both licensed and unlicensed bands. Here, the drones are the agents, and their actions pertain to choosing the proper duty cycle to enable a co-existence between cellular and WiFi communications via the drones \cite{rein1}. Therefore, the key step in applying RL in IoT scenarios is finding appropriate counterparts to each component of RL.\\ 
\indent The state transition function is the most challenging to define in the IoT, because it is typically required for all devices to know this function. In the given example of power management, the state transition function is defined such that the next state only depends on the current action \cite{learning2}. However, for other resource management applications in the IoT, the state transition function may be complex and dependent on other devices as well. For instance, in radio resource management, the actions may be choosing a channel to use for uplink communication, and the state can be signal-to-interference-plus-noise ratio of the selected channel \cite{learning8}. In this application, the state transition function depends on other devices, and the next state is uncertain as the actions of other devices are unknown. Although the use of memory improves the performance of RL by allowing the uncertain states to exist \cite{fmrl}, it is necessary to investigate the tradeoffs associated with having a memory.\\
\begin{table*}[t]
\centering
\caption{Summary of the Proposed Learning Frameworks}
\label{learning_table}
\begin{tabular}{|c|l|l|l|}
\hline
\textbf{Learning frameworks} & \multicolumn{1}{c|}{\textbf{Advantages}} & \multicolumn{1}{c|}{\textbf{Limitations}} & \multicolumn{1}{c|}{\textbf{Applications}} \\ \hline
Machine Learning             & \begin{tabular}[c]{@{}l@{}} \textbullet \ Diversity of available techniques \\ with a multitude of applications. \\ \textbullet \ Useful in processing the massive \\ data collected in IoT to reduce \\ resource consumption.  \\ \textbullet Enabler for big data analytics \\ and predictive solutions. \end{tabular}                                          & \begin{tabular}[c]{@{}l@{}} \textbullet \ Implementation typically centralized. \\ \textbullet \ Need for significant computational \\ capabilities. \\ \textbullet \ Need for extensive training data.\end{tabular}                                & \begin{tabular}[c]{@{}l@{}} \textbullet \ Data aggregation and \\ compression \cite{pca1, pca1e}. \\ \textbullet \ Query processing \cite{bigdata2}. \\ \textbullet \ Big data analytics \cite{bigdata, ml1, learning1}. \\ \textbullet \ IoT security \cite{security1, security2}. \end{tabular} \\ \hline
Sequential Learning          & \begin{tabular}[c]{@{}l@{}} \textbullet \ Distributed implementation. \\ \textbullet \ Flexible memory and resource \\ requirements. \\ \textbullet \ Lack of a need for extensive \\ knowledge of the system. \\ \textbullet \ Ability to effectively learn \\ unknown parameters.\end{tabular}                                                                                         & \begin{tabular}[c]{@{}l@{}} \textbullet \ Reliance on M2M communication. \\ \textbullet \ Limited to binary state learning. \\ \textbullet \ Need for a private signal for learning. \end{tabular}                                              & \begin{tabular}[c]{@{}l@{}} \textbullet \ Enhanced event detection. \\ \textbullet \ Dynamic resource management \\ under uncertainty \cite{tae}. \\\textbullet \ Network operation adaptation. \end{tabular} \\ \hline
Reinforcement Learning       & \begin{tabular}[c]{@{}l@{}} \textbullet \ Distributed implementation.\\ \textbullet \ Asynchronous operation. \\ \textbullet \ Low computational complexity. \\ \textbullet \ Ability to find game-theoretic \\ equilibrium solutions. \end{tabular}                                                      & \begin{tabular}[c]{@{}l@{}} \textbullet \ Significant overhead to reach steady \\ state. \\ \textbullet \ Need for complete information about \\ state transition. \\ \textbullet \ Higher computational complexity \\ with incomplete information. \end{tabular}                                     & \begin{tabular}[c]{@{}l@{}}\textbullet \ Power control \cite{learning2}. \\ \textbullet \ Radio resource management for \\ IoT environments \cite{learning8, rein2, rein3}. \\ \textbullet \ Dynamic scheduling for energy \\ efficiency \cite{rein2}. \\ \textbullet \ Use of drones for enhanced \\ communications \cite{rein1}. \end{tabular} \\ \hline
\end{tabular} \label{tab:learning}
\end{table*}
\indent There are different types of memory that RL algorithms can adapt to mitigate the effects of having unknown states \cite{fmrl}. Moreover, the different approaches of using memory require different amounts of information and introduce varying levels of computational complexity \cite{fmrl}. Furthermore, different approaches have different properties in terms of scalability, convergence time, and quality of steady state \cite{fmrl}. Therefore, similar to SL, it is necessary to choose a type of memory for RL that is most suitable considering the available resources and QoS requirements. This will make RL much more applicable for the IoT.\\
\indent In summary, the IoT provides an environment that is ripe for applying a variety of learning algorithms. However, such algorithms must be properly designed and tailored to the intrinsic properties of the IoT. The various challenges, opportunities, and applications of learning are summarized in Table \ref{tab:learning}. Next, we discuss how learning approaches can be designed in a way to cope with the heterogeneity of the IoT environment.
%%%%%%%%%%%%%%%%%%%%%%%%%%%%%%%%%%%%%%%%%%%%%%%%%%%%%%%%%%%%%%%%%%%%%%%%%%%%%%%%%%%%
\section{Managing IoT Heterogeneity using Cognitive Hierarchy Theory}%%*** MORE LINK TO BEFORE, BETTER FLOW TO HERE***
\subsection{Motivation}
\indent The aforementioned learning frameworks exhibit different resource requirements and can achieve varying levels of accuracy in the learned parameters. For instance, the agents using SL can have more observations of previous agents in the sequence with more memory, thus having more information about the underlying state \cite{tae}. Therefore, SL schemes with more memory will allow the IoT devices to more accurately learn the underlying state, thus achieving lower probability of estimation error. On the other hand, the IoT devices can also have different available resources. For example, smartphones will be able to do advanced learning methods as they are computationally capable and have extensive memory, while simple sensors with limited resources available must resort to elementary learning methods. Therefore, the difference in the available resources among the IoT devices will cause them to reach different learning outcomes. Although most existing learning frameworks typically assume that the agents have the same available resources, it is necessary to capture and exploit the heterogeneity in available resources among the IoT devices.\\
\indent In this regard, \emph{cognitive hierarchy theory (CHT)} \cite{Camerer_acognitive, cht1}, and \cite{cht2}, is a promising tool to accurately model the IoT, because it provides a modeling framework that can properly capture the heterogeneity among the agents. CHT divides a number of agents into different groups of varying rationality levels. Different rationality levels can be interpreted as different levels of available resources among the IoT devices. Thus, CHT will allow each group of agents (devices) at a given rationality level to choose a learning framework that is most appropriate based on their available resources. Therefore, CHT will make it possible to integrate different learning frameworks at different levels in the IoT, thus maximizing the overall usage of the resources and providing a more realistic model for a heterogeneous IoT system.
\subsection{Preliminaries and Key Concepts}
\indent CHT is a branch of behavioral game theory and is based on the concept of bounded rationality \cite{cht1, cht2}. In general, bounded rationality means that each agent finds the best strategy based on its accessible information, its computational capacity, and the time available. In CHT, it is assumed that the agents are distributed into discrete levels of rationality. Agents belonging to the lowest level 0 are completely irrational and choose their strategy randomly. Agents at any higher level $k \geq 1$ believe that all others belong to the levels lower than $k$ and choose their optimal strategy based on their beliefs. Therefore, to find the optimal strategy, any agent at level $k \geq 1$ starts by computing the strategy of level 1 agents then level 2 agents up to level $k-1$ agents and finally computes its optimal strategy, hence performing $k$ levels of thinking.\\
\indent The most popular CHT model is the Poisson model \cite{cht2}. It relies on the following considerations. First, it considers that the agents are distributed into the rationality levels according to a Poisson distribution $f$ with rate $\tau$. The Poisson distribution has been shown in \cite{Camerer_acognitive} to be a good model for the situations in which, as the rationality level $k$ grows larger, fewer agents will be at a higher level than $k$. The second assumption is that level 0 agents choose their strategies randomly according to a uniform distribution. This is commonly known as the overconfidence assumption. The last assumption is that each agent at level $k \geq 1$ knows the true proportions $f(0), f(1), ..., f(k-1)$ of agents at lower rationality levels.\\
\indent CHT is useful for the IoT because the different rationality levels of CHT can correspond to the different resources available for the IoT devices. Hence, by grouping the devices into the rationality levels based on their available resources, the CHT framework allows each IoT device to find its optimal strategy based on its own, individual computational capability. Note that the CHT model is different from classical hierarchical approaches, such as Stackelberg game and its variants. In such approaches, it is assumed that all agents are fully rational and that the hierarchy is defined based on the roles of the agents (leaders vs. followers) and not in terms of the rationality.\\
\indent The concept of CHT can be further extended beyond the existing models, such as \cite{Camerer_acognitive} and \cite{cht1}, to provide a learning scheme for the heterogeneous IoT devices. In this case, the devices at each rationality level can adapt a learning framework that matches their available resources and application. For example, a data aggregator will use ML to compress data, resource-constrained sensors will use finite memory SL with small memory for event detection, and smartphones will used sophisticated RL with relatively large memory for resource management. Moreover, resource-constrained sensors can use simple RL with small memory as they are low rationality, while smartphones would use SL with big memory as they are high rationality. Therefore, the memory size and the computational complexity associated with learning will depend on the rationality level of the device. Applying a hierarchical approach to learning was previously considered in \cite{hierarchy_example}, however, this approach cannot capture heterogeneity as it assumed that all the agents use the same learning framework.
\subsection{Applications of CHT for the IoT}
\begin{figure}[t]
	\centering
	\includegraphics[width=8.5 cm,height=6.5cm,angle=0]{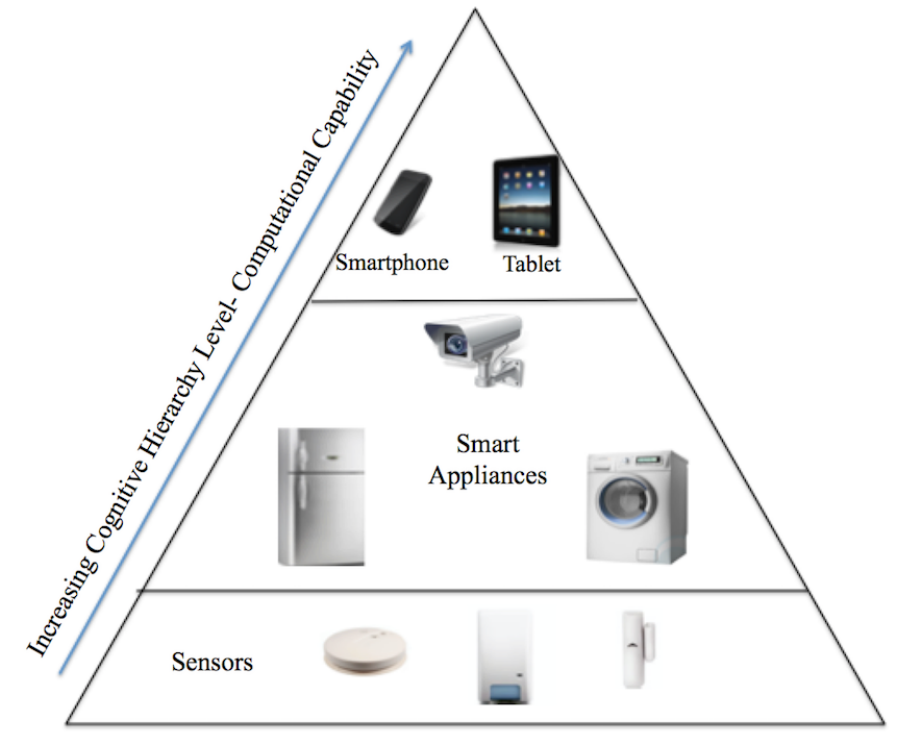}
	\caption{Distribution of IoT devices in different CHT levels.}
	\vspace{-0.4 cm}\label{fig:iotmodel}
\end{figure}
\indent CHT can be applied to IoT problems that involve distributed optimization at the IoT devices, such as distributed multiple access, resource allocation, and secure transmissions. For example, we have proposed a CHT based approach in \cite{CHIoT} to address the problem of distributed uplink multiple access in an IoT network composed of MTDs and HTDs. The MTDs have different characteristics, such as packet sizes, transmission powers, or queue buffer size. In this model, a random access is used for multiple access. In this scheme, a collision occurs if two or more devices transmit simultaneously. Furthermore, the QoS requirement of each device is defined depending on its type. For many IoT devices, such as health care sensors and alarm systems, it is necessary to deliver each packet within a strict deadline while minimizing their energy consumption. HTDs, on the other hand, are more interested in maximizing their transmission rates while keeping their energy consumed within a certain budget. Also due to collisions, the choice of transmission probability of each device affects the performance of all other devices, which motivates to use the game theoretic approaches.\\
\indent Further, since the IoT devices have different resources available, different rationality levels are used to correspond to the heterogeneous resource availabilities of the IoT devices. A device at rationality level $k$ does more thinking than devices at lower levels and hence should be more computationally capable. Fig. \ref{fig:iotmodel} shows an example on how some IoT devices can be distributed according to CHT rationality levels.\\
\indent Here, the Poisson model is chosen as it is suitable to model the distribution of devices in the IoT network. This is because there is a limit on the computational capability of the devices especially in terms of memory and processing power. In fact, devices with higher computational capability are more expensive and hence are fewer in number. The second assumption of the Poisson model related to level-0 devices also holds in the IoT network for resource-constrained devices (e.g., MTDs), whose most feasible strategy is to pick randomly. However, in the approach that we proposed in \cite{CHIoT}, the overconfidence assumption is modified so that an agent at rationality level $k$ is aware of the presence of agents at the same rationality level. The overconfidence assumption is suitable for games in which the agents are humans, however, in the studied IoT scenario, the agents are IoT devices that can observe other devices of the same computational capabilities.

\begin{figure}[t]
	\centering
	\includegraphics[width=8.5 cm,height=6.5cm,angle=0]{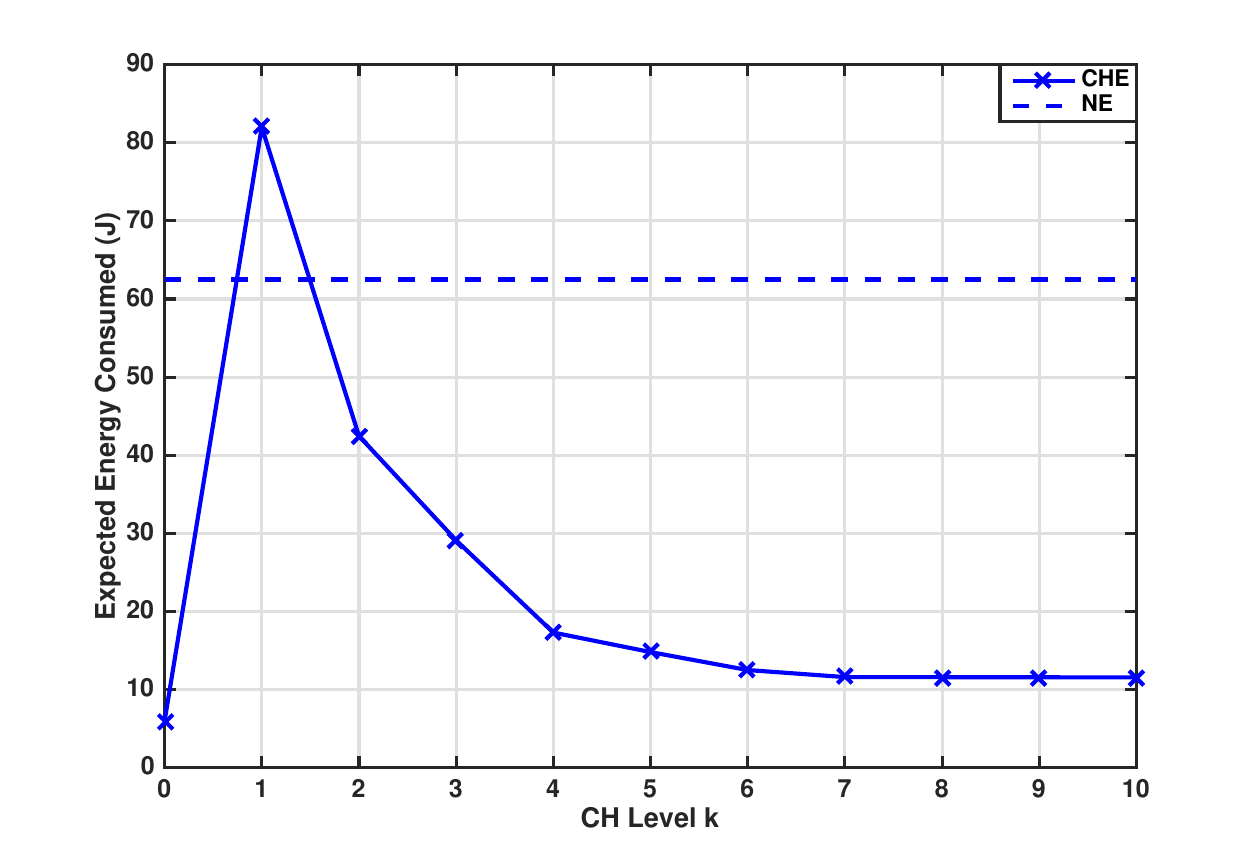}
	\caption{Expected energy consumed at the cognitive hierarchy equilibrium (CHE) solution vs. CHT level.
	}\vspace{-0.4 cm}\label{fig:chperformance}
\end{figure}

\indent Using this CHT model, in \cite{CHIoT}, we have shown that any device at level $k$ needs to solve a nonlinear, nonconvex optimization problem to find its optimal strategy. Also, the strategy of the device at rationality level $k$ is dependent on the strategies of agents at lower rationality levels. Hence, a device at level $k$ needs to solve $k$ nonlinear optimization problems to find its optimal strategy. This result illustrates the connection between the CHT rationality levels and the computational capabilities of the MTDs as an IoT device at level $k$ needs to do more computations than the devices at lower levels. It is further shown that the proposed CHT approach brings considerable performance improvements. Fig. \ref{fig:chperformance} shows that starting from CHT level 1, the energy consumed by an MTD decreases as its CHT level increases and becomes considerably lower than the energy consumed resulting from the classical Nash equilibrium (NE) approach, which shows that the proposed CHT approach provides a fair tradeoff between the achieved performance and the computational capability of the device.\\
\indent Clearly, CHT is a promising modeling framework within which one can naturally capture the heterogeneity of the IoT and integrate different learning approaches that are directly mapped to the resource limitations of the IoT devices.
\section{Summary}
\indent In this article, we have provided a comprehensive overview on the use of a variety of learning techniques within a wide range of applications in the IoT. First, we have discussed the general properties needed for learning techniques to be applied and developed for the purpose of IoT optimization and resource management. Then, we have reviewed the advantages and the limitations of three popular learning frameworks: machine learning, sequential learning, and reinforcement learning. For each such framework, we have identified the inherent properties, such as computational complexity, memory requirements, and learning performance. Furthermore, we have introduced the fundamentals of cognitive hierarchy theory, which can be used to categorize the IoT devices into different levels of rationality, depending on their capabilities, thus providing a more realistic model of the IoT with heterogeneous devices. For each level of rationality, CHT makes it possible to determine which learning frameworks are most suitable given the requirements for learning and the available resources of the IoT devices in that level of cognitive hierarchy. In summary, this article provides a comprehensive reference on developing and applying several classes of learning techniques within well-defined IoT applications.

\bibliographystyle{IEEEtran}
\bibliography{references}
\end{document}